\documentclass[sigconf]{acmart}

\usepackage{multicol}
\usepackage{multirow}
\usepackage{amsmath}
\usepackage{balance}

\DeclareMathOperator*{\argmin}{arg\,min}
\AtBeginDocument{%
  \providecommand\BibTeX{{%
    \normalfont B\kern-0.5em{\scshape i\kern-0.25em b}\kern-0.8em\TeX}}}

\copyrightyear{2021} 
\acmYear{2021} 
\setcopyright{acmcopyright}\acmConference[ICTIR '21]{Proceedings of the 2021 ACM SIGIR International Conference on the Theory of Information Retrieval}{July 11, 2021}{Virtual Event, Canada}
\acmBooktitle{Proceedings of the 2021 ACM SIGIR International Conference on the Theory of Information Retrieval (ICTIR '21), July 11, 2021, Virtual Event, Canada}
\acmPrice{15.00}
\acmDOI{10.1145/3471158.3472232}
\acmISBN{978-1-4503-8611-1/21/07}

\begin{document}
    \fancyhead{}
\title{Asking Clarifying Questions Based on Negative Feedback in Conversational Search}

	\author{Keping Bi}
	\affiliation{
		\institution{University of Massachusetts Amherst}
		\city{Amherst} 
		\state{MA} 
		\country{USA}
	}
	\email{kbi@cs.umass.edu}

	\author{Qingyao Ai}
	\affiliation{
		\institution{University of Utah}
		\city{Salt Lake City} 
		\state{UT} 
		\country{USA}
	}
	\email{aiqy@cs.utah.edu}

	\author{W. Bruce Croft}
	\affiliation{%
		\institution{University of Massachusetts Amherst}
		\city{Amherst} 
		\state{MA} 
		\country{USA}
	}
	\email{croft@cs.umass.edu}

\begin{abstract}
  Users often need to look through multiple search result pages or reformulate queries when they have complex information-seeking needs. Conversational search systems make it possible to improve user satisfaction by asking questions to clarify users' search intents. This, however, can take significant effort to answer a series of questions starting with ``what/why/how''. 
  To quickly identify user intent and reduce effort during interactions, we propose an intent clarification task based on yes/no questions where the system needs to ask the correct question about intents within the fewest conversation turns. In this task, it is essential to use negative feedback about the previous questions in the conversation history. To this end, we propose a Maximum-Marginal-Relevance (MMR) based BERT model (MMR-BERT) to leverage negative feedback based on the MMR principle for the next clarifying question selection. Experiments on the Qulac dataset show that MMR-BERT outperforms state-of-the-art baselines significantly on the intent identification task and the selected questions also achieve significantly better performance in the associated document retrieval tasks. 

\end{abstract}




\maketitle

\section{Introduction}
\label{sec:introduction}
In traditional Web search, users with complex information needs often need to look through multiple pages or reformulate queries to find their target information. In recent years, intelligent assistants such as Google Now, Apple Siri, or Microsoft Cortana make it possible for the system to interact with users through conversations. By asking questions to clarify ambiguous, faceted, or incomplete queries, 
conversational search systems could improve user satisfaction with better search quality. Thus, how to ask clarifying questions has become an important research topic. 

There are two typical types of clarifying questions: \textit{special questions} beginning with what/why/how etc. and \textit{general (yes/no) questions} that can be answered with ``yes'' or ``no''. Special questions often let a user give specific information about a query such as ``What do you want to know about COVID-19?'' for the user query ``COVID-19''. This kind of question is usually more difficult and requires more user effort to answer than questions such as ``Do you want to know the symptoms of COVID-19?'' With an explicit option in the question, users can easily confirm or deny by saying ``yes'' or ``no''.
In addition to requiring less effort from users, yes/no clarifying questions make it easier for the system to decide when to show text retrieval results. Users' affirmative answers could enhance the system's confidence in the text retrieval performance. 
\begin{figure}
	\includegraphics[width=0.35\textwidth]{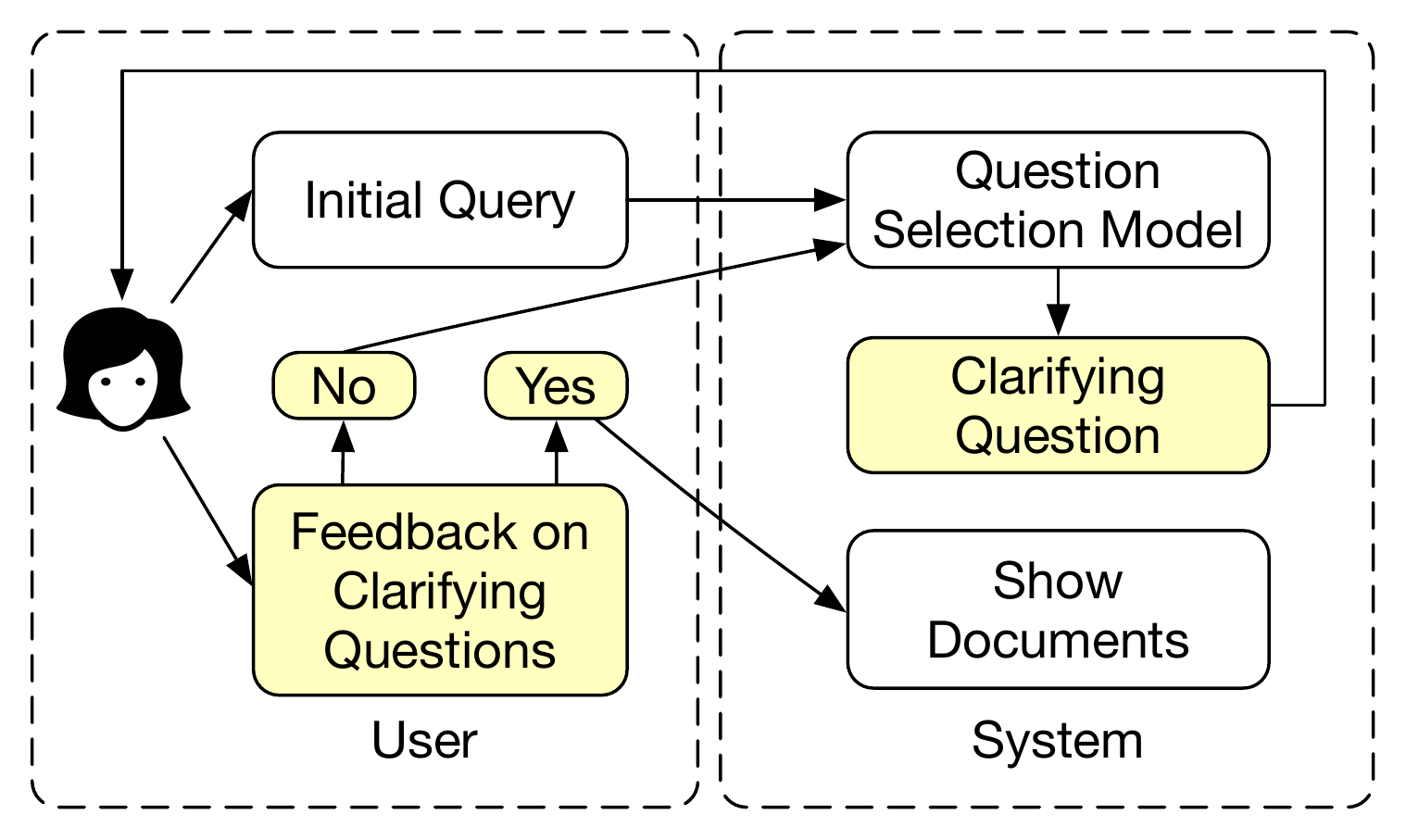} %
	\caption{A workflow of the intent clarification task. }
	\label{fig:neg_cq_conv_flow}
\end{figure}

Given these observations, we propose an intent clarification task based on yes/no questions where the target of the system is to select the correct questions about user intent within the fewest conversation turns, shown in Figure \ref{fig:neg_cq_conv_flow}. After the user issues an initial query, the system asks yes/no clarifying questions to the user. When the user provides negative feedback, the system asks another question to confirm the user's intent. 
When the intent is confirmed or the limit of conversation turns is reached\footnote{Because it is impractical to ask unlimited number of questions to users, it is common for conversational search systems to set a limit to the number of asked questions.}, the system returns the results of document retrieval. 
In the intent clarification task, it is essential to leverage negative feedback about the previously asked questions in the conversation history effectively to select the next question. The principle of using negative feedback is to find a candidate that is dissimilar to the negative results while keeping it relevant to the query. In Web search, documents with negative judgments have limited impact on identifying relevant results due to the large number of potential non-relevant results \cite{wang2007improve, wang2008study, karimzadehgan2011improving}. In contrast, the intent space of a query is much smaller, providing more opportunity to leverage negative feedback from previous clarifying questions. 

In this paper, we train an initial model to select the first clarifying question based on the original query.
Then we propose a maximum-marginal-relevance (MMR) based BERT model (MMR-BERT) to leverage negative feedback in the conversation history for the next clarifying question selection. 
Experiments on the Qulac \cite{aliannejadi2019asking} dataset show that MMR-BERT outperforms the state-of-the-art baselines significantly on the intent clarification task and the selected questions also achieve significantly better performance in the associated document retrieval tasks. 
We then give a detailed analysis of each method's number of success conversations, the impact of topic/facet type on each model, and the success/failure cases of our model compared to the best baseline. 
\section{Related Work}
There are three threads of work related to our study: conversational search and question answering (QA), asking clarifying questions, and negative feedback. 

\textbf{Conversational Search and QA.}
The concept of information retrieval (IR) through man-machine dialog dates back to 1977 \cite{oddy1977information}. Other early work in conversational IR includes an intelligent intermediary for IR, named as I$^3$R, proposed by \citet{croft1987i3r} in 1987, and an interactive IR system using script-based information-seeking dialogues, MERIT, built by \cite{belkin1995cases} in 1995. 
In recent years, task-based conversational search based on natural dialogues has drawn much attention. \citet{radlinski2017theoretical} proposed a theoretical framework for conversational IR. \citet{vtyurina2017exploring} studied how users behave when interacting with a human expert, a commercial intelligent assistant, and a human disguised as an automatic system.  \citet{spina2017extracting} studied how to extract audio summaries for spoken document search. \citet{trippas2018informing} suggested building conversational search systems based on the commonly-used interactions from human communication. 
Most recently, \citet{yang2018response} conducted response ranking based on external knowledge given a conversation history. \citet{wang2021controlling} propose to control the risk of asking non-relevant questions by deciding whether to ask questions or show results in a conversation turn.

Conversational question answering defines the task of finding an answer span in a given passage based on the question and answers in the conversation history such as CoQA \cite{reddy2019coqa} and QuAC \cite{choi2018quac}. \citet{qu2020open} extended the task by introducing a step of retrieving candidate passages for identifying answer span. This is more practical in real scenarios where ground truth passages that contain the answers are often unavailable.

In this paper, we focus on the next clarifying question selection based on negative feedback to identify users' true intent in the fewest conversation turns, which differs from most existing work in conversational search. Also, our intent clarification task is fundamentally different from the objective of conversational QA.


\textbf{Asking Clarifying Questions.}
In the TREC 2004 HARD track \cite{allan2005hard}, systems can ask searchers clarification questions such as whether some titles seem relevant to improve the accuracy of IR. 
\citet{rao2018learning} collected a clarifying question dataset from the posts in StackOverflow and proposed to select clarification questions based on the expected value of perfect information considering the usefulness of potential answers to a candidate question. Later, \citet{rao2019answer} extended the work by using the utility \cite{rao2018learning} in a reinforcement learning framework in product QA to handle cases where contexts such as product information and historical questions and answers are available. 
\citet{zhang2018towards,sun2018conversational} proposed to ask users questions about their preferred values on aspects of a product for conversational product search and recommendation.
\citet{wang2018learning} observed that a good question is often composed of interrogatives, topic words, and ordinary words and devised typed encoders to consider word types when generating questions. 
\citet{cho2019generating} proposed a task of generating common questions from multiple documents for ambiguous user queries. \citet{xu2019asking} studied whether a question needs clarification and introduced a coarse-to-fine model for clarification question generation in knowledge-based QA systems. \citet{zamani2020generating} extracted the facets of a query from query logs and generated clarifying questions through template or reinforcement learning with weak supervision. 

To study how to ask clarifying questions in information-seeking conversations, \citet{aliannejadi2019asking} collected clarifying questions through crowd-sourcing in a dataset called Qualc based on the ambiguous or faceted topics in the TREC Web track \cite{clarke2009overview,clarke2012overview}. They proposed to select the next clarifying question based on BERT representations and query performance prediction. 
Later, \cite{hashemi2020guided} extended the idea of pseudo relevance feedback and leveraged top-retrieved clarifying questions and documents for document retrieval and next clarifying question selection on Qulac. 
\citet{aliannejadi2020convai3} then organized a challenge on clarifying questions for dialogue systems that raises the questions on when to ask clarifying questions during dialogues and how to generate the clarifying questions.

Most existing work evaluates models based on either the initial query or pre-defined conversation history, i.e., the models always select the next question based on static conversation turns instead of its previously selected questions. In contrast, we select the next questions dynamically considering previous questions, which is more practical. 
Also, other studies do not differentiate responses that are confirmation or denial. In contrast, we address how to leverage negative feedback in the response. 


\textbf{Negative Feedback.}
Existing work on negative feedback has been relatively sparse and mostly focuses on document retrieval for difficult queries. \citet{wang2007improve} proposed to extract a negative topic model from non-relevant documents from its mixture with the language model of the background corpus. 
The Rocchio model \cite{rocchio1971relevance} considers both positive and negative feedback and can be used when only negative feedback is available. 
\citet{wang2008study} compared various negative feedback methods in the framework of language model or vector space model. Later, \cite{karimzadehgan2011improving} proposed a more general negative topic model that further improved the performance of difficult queries. \citet{peltonen2017negative} designed a novel search interface where users can provide feedback on the keywords of non-relevant results. 

Negative feedback has also been studied in recommendation and product search. \citet{zagheli2017negative} proposed a language model based method to avoid recommending texts similar to documents users dislike. \citet{zhao2018recommendations} considered skipped items as negative feedback and used it together with positive feedback to recommend items by trial and error. 
\citet{bi2019conversational} leveraged user feedback on finer-grained aspect-value pairs extracted from non-relevant results in conversational product search. 

Unlike these studies, we study how to leverage negative feedback to clarifying questions that are much shorter than documents in open-domain information-seeking conversations. Our model is based on pre-trained BERT \cite{devlin2018bert} models and the Max Marginal Relevance (MMR) \cite{carbonell1998use} principle. 

\section{Conversation Intent Clarification}
\label{sec:methods}
In this section, we first introduce the definition of the conversation intent clarification task. To approach the task, we propose a two-step method to ask clarifying questions in the conversation. We illustrate the model for initial clarifying question selection in Section \ref{subsec:init_retrieval} and the model that selects the next question using negative feedback to previous questions in Section \ref{subsec:cq_negfeedback}. 

\begin{figure}
	\includegraphics[width=0.5\textwidth]{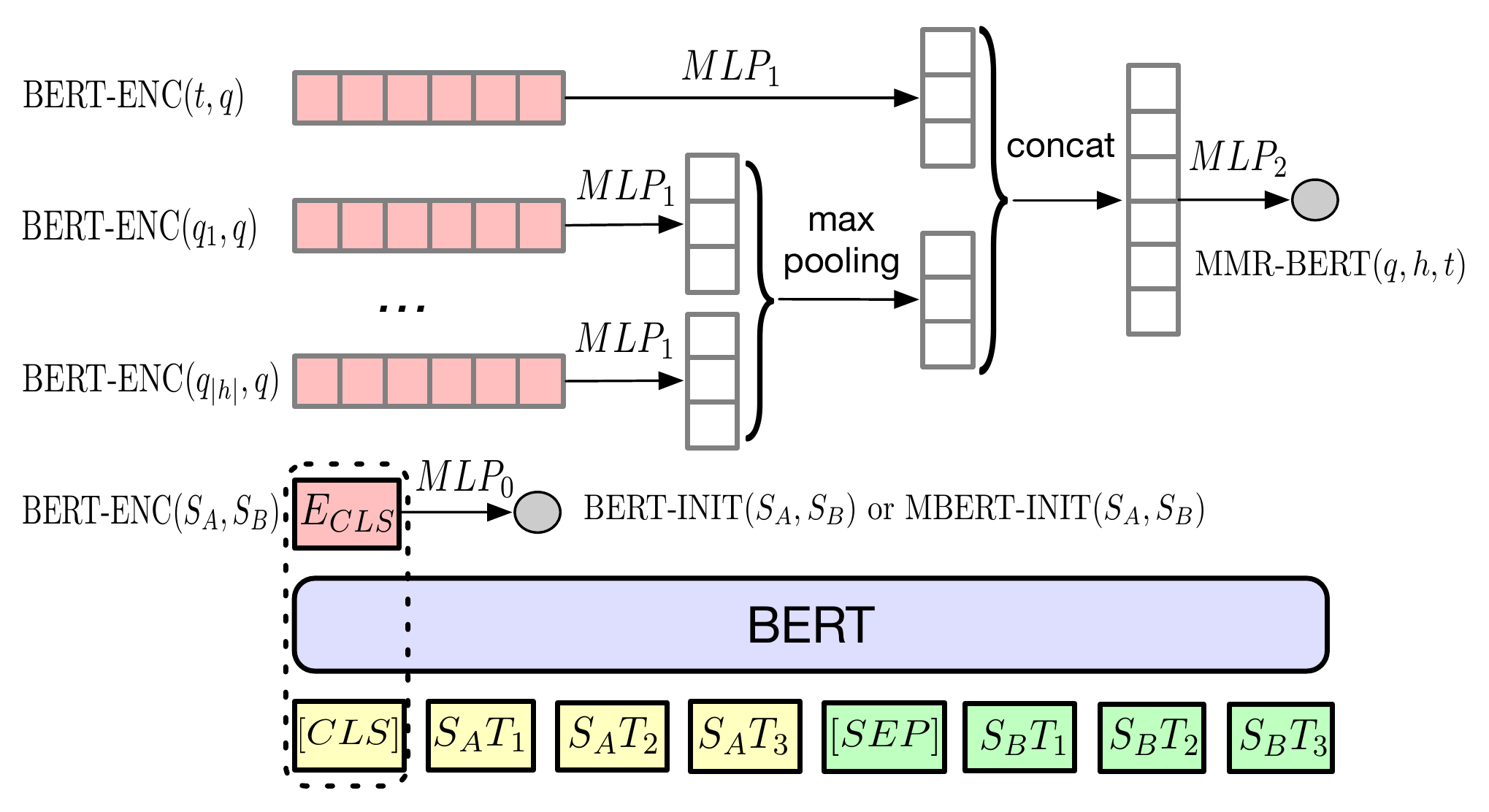} %
	\caption{Our Maximal Marginal Relevance based BERT Model (MMR-BERT). }
	\label{fig:mmr-bert}
\end{figure}
\subsection{Task Formulation}
Suppose that a user has a specific information need about an ambiguous or faceted topic $t$. The user issues $t$ as a query to the system \footnote{We use topic and query interchangeably in the paper}. Let $h=((q_1, a_1), (q_2, a_2), \cdots, (q_{|h|}, a_{|h|}))$ be the conversation history between the user and the system, where the system asks the user $|h|$ clarifying questions $Q_h = \{q_i | 1\leq i \leq |h|\}$ about the potential intents behind the topic, and the user confirms or denies the corresponding intent indicated in $q_i$ with $a_i$. For any candidate question $q$, its label $y(q)=2$ if it covers the user's true intent, $y(q)=1$ if it covers other intents of $t$, and $y(q)=0$ if it is not relevant to $t$. The system's target is to identify the user's true intent within the fewest interactions, i.e., $\argmin(|Q^{\bigstar}=\{q|y(q)=2|)$.
Since it is not practical to ask too many questions, the system ends the conversation and returns the document retrieval results whenever the user's intent is confirmed or the limit of conversations turns $k$ ($|h|\leq k$) is reached. 

\subsection{First Clarifying Question Selection}
\label{subsec:init_retrieval}
The first clarifying question is especially important to elicit user interactions as it will impact the effectiveness of all the future questions and user interactions. 
The information available to select the initial question is the query itself. Thus it is essential to effectively measure the relevance of a candidate question by how it matches the user query.

\textbf{Query-question Matching.}
In recent years, BERT \cite{devlin2018bert} has shown impressive performance in short-text matching tasks by pre-training contextual language models with large external collections and fine-tuning the model based on a local corpus. 
We leverage BERT to select questions in the intent clarification task. Specifically, we select the first question based on the relevance score of matching a candidate $q$ to topic $t$ calculated with BERT:
\begin{equation}
\label{eq:init_bert}
s(q,t) = MLP_0(\text{BERT-ENC}(q,t))
\end{equation}
where BERT-ENC$(S_A, S_B)$ is the output vector of matching sentence A ($S_A$) and sentence B ($S_B$) as shown in Figure \ref{fig:mmr-bert}, $MLP_0$ is a multilayer perceptron (MLP) with output dimension 1. Specifically, BERT-ENC$(S_A, S_B)$ inputs the token, segment, and position embeddings of the sequence (\textsl{[CLS], tokens in $S_A$, [SEP], tokens in $S_B$}) to the pre-trained BERT model \cite{devlin2018bert} and take the vector of [CLS] after the transformer encoder layers as output. 

\textbf{Loss Function.}
We have two ways of calculating the training loss. 
As a first option, assuming that we do not have any prior knowledge about each user's intent, the retrieval of the first question should simply focus on retrieving questions that are relevant to the initial query string $t$. 
Thus we collect a set of query pairs $Q^P$ and each pair consists of a relevant and a non-relevant question, i.e., $Q^P=\{(q^+, q^-)| y(q^+)>0, y(q^-)=0 \}$. We consider all the questions with positive labels having the same label 1, i.e., $y'(q) = \mathcal{I}(y(q) > 0)$, where $\mathcal{I}$ is an indicator function and equals to 1 when the input condition is true otherwise it is 0.
The probability of question $q$ in the entry (pair) $E$ ($E\in Q^P$) being relevant to query topic $t$ is calculated with the softmax function:

\begin{equation}
\label{eq:prob}
   Prob(y'(q)=1) = \frac{\exp(s(q,t))}{\sum_{q'\in E}\exp(s(q',t))}, E \in Q^P.
\end{equation}
Then the loss function $\mathcal{L}$ is the cross-entropy between the binary question labels $(1,0)$ of the pair and the probability distribution of $(Prob(y'(q^+)=1), Prob(y'(q^-)=1))$:

\begin{equation}
\label{eq:pair_loss}
    \mathcal{L}_{\text{BERT-INIT}} = -\sum_{E \in Q^P} \sum_{q \in E} y'(q) \log Prob(y'(q)=1).
\end{equation}
In this case, the loss function is essentially pairwise loss. We refer to the model trained with $Q^P$ as \textit{BERT-INIT}. 

Among the relevant questions of the same query, only questions that match user intents can receive positive feedback and have label 2. As a second option, when we further consider which relevant questions are more likely to receive positive feedback in a prior distribution, the multi-grade label of a question can be used for training. 
We extend the set of question pairs $Q^P$ to question triplets $Q^T=\{(q^{\bigstar}, q^{*}, q^-)| y(q^{\bigstar})=2, y(q^{*})=1, y(q^-=0)\}$ and still use the cross-entropy loss to optimize the model. In other words, we train the model according to:

\begin{equation}
\label{eq:triplet_loss}
    \mathcal{L}_{\text{MBERT-INIT}} = -\sum_{E \in Q^T} \sum_{q \in E} y(q) \log Prob(y(q)>0),
\end{equation}
where $Prob(y(q)>0)$ is calculated based on Equation \eqref{eq:prob} with $Q^P$ replaced by $Q^T$ and $E$ is an entry of triplet.
As in \cite{ai2018learning}, this loss function can be considered as a list-wise loss of the constructed triplets. Since the probability of each question to be a target question is normalized by the scores of all the three questions in the triplet, maximizing the score of question with label 2 will reduce the score of questions with label 1 and 0. Also, questions with larger labels have more impact to the loss. This ensures that the model is optimized to learn higher scores for questions that have larger labels. We refer to this model as \textit{MBERT-INIT}.

\subsection{Clarifying Intents Using Negative Feedback}
\label{subsec:cq_negfeedback}
While the only basis of the system's decision is topic $t$ in the first conversation turn, the system can refer to conversation history in the following interactions. As we assume that the system will terminate the conversation and return the documents when the user confirms the question with positive feedback, all the available information for selecting the next clarifying question besides the topic $t$ is negative feedback. It means that the next question should cover a different intent from previous questions while being relevant to topic $t$. 


Inspired by the maximal marginal relevance (MMR) principle in search diversification studies \cite{carbonell1998use}, here we propose an MMR-based BERT model (MMR-BERT) to leverage negative feedback in the conversations. 
In search diversification, the basic idea of MMR is to select the next document by maximizing its relevance to the initial query and dissimilarities to previously selected documents.
Similarly, in MMR-BERT, we select the next question by jointly considering the relevance of each candidate question with respect to the initial topic $t$ and their similarities to previous questions.
Let $Q$ be the question candidate set, and $Q_h=\{q_i|1\leq i \leq |h|\}$ be the set of questions in the conversation history $h$.
Let BERT-ENC$(S_A,S_B)$ be a matching function that takes two pieces of text (i.e., $S_A$ and $S_B$) as input and outputs an embedding/feature vector to model their similarities. \footnote{Here we use BERT encoder as our matching model because it has been shown to be effective in modeling the latent semantics of text data, which is important for our task since different facets of the same topic often have subtle semantic differences that cannot be captured by simple methods such as keyword matching.}
As shown in Figure \ref{fig:mmr-bert}, MMR-BERT first obtains the matching of the topic $t$ with candidate question $q$, i.e., BERT-ENC$(t,q)$ and the matching between each previous question $q_i (1\leq i \leq |h|$) and $q$, i.e., BERT-ENC$(q_i, q)$. Then it maps the obtained vectors to lower d-dimension space ($\mathbb{R}^d$) with a multilayer perceptron (MLP) $MLP_1$, where each layer is a feed-forward neural network followed by Rectified Linear Unit (ReLU) activation function. The parameters in $MLP_1$ are shared across multiple matching pairs to let the condensed vectors comparable. Formally, the final matching between $x$ and $q$ is:
\begin{equation}
\label{eq:matching}
\begin{split}
& o(x,q) = MLP_1(\text{BERT-ENC}(x,q)) \in \mathbb{R}^{d} \\
& x=t \text{ or } q_i, 1\leq i \leq |h|
\end{split}
\end{equation}
The final score of $q$ is computed as: 
\begin{equation}
\label{eq:mmr-bert}
\!\! \text{MMR-BERT}(q,t,h) \! = \! MLP_2([o(t,q);\!MaxPool_{1\leq i \leq |h|}o(q_i, q)])\!\!
\end{equation}
where $MaxPool$ represents apply max pooling on a group of vectors, $[\cdot;\cdot]$ denotes the concatenation between two vectors, $MLP_2$ is another MLP for projection to $\mathbb{R}^1$. 

Given the user's negative feedback to the asked questions in the conversation history $h$, the probability of a candidate $q$ covering user intent is calculated according to:
\begin{equation}
\label{eq:hist_prob}
    Prob(y(q)\!=\!2|h) \!= \! \frac{\exp(\text{MMR-BERT}(q,t,h))}{\sum_{q'\in E}\exp(\text{MMR-BERT}(q',t,h))}, E \!\in\! Q^T \!,
\end{equation}
where $Q^T$ is a set of triplets, $E$ is a triplet of questions with label 2, 1, and 0, as in Section \ref{subsec:init_retrieval}. 
To differentiate the questions that would receive positive feedback from users and questions that are relevant to the topic $t$ but do not match user intents, we use the multiple-grade labels in the loss function, as MBERT-INIT in Section \ref{subsec:init_retrieval}. Since $Prob(y(q)=2,h)=Prob(y(q)=2|h)Prob(h)$ and $Prob(h)$ is fixed for topic $t$ during training. The loss function is: 
\begin{equation}
\label{eq:mmrbert_loss}
    \mathcal{L}_{\text{MMR-BERT}} \propto -\sum_{E \in Q^T} \sum_{h \in H(E)} \sum_{q \in E} y(q) \log Prob(y(q)=2|h),
\end{equation}
where $H(E)$ is the history set of conversation turns of length 0, 1, 2, and so on, corresponding to triplet entry $E$. For example, if the questions $q_a$,$q_b$, and $q_c$ are already asked for topic $t$, $H(E)=\{\emptyset, \{q_a\}, \{q_a,q_b\}, \{q_a,q_b,q_c\}\}$. The answers in the history are omitted in the notation since they are all ``no".
In this way, questions that cover similar intents to historically asked questions $Q_h$ have lower labels than the questions that have target intents and thus will be punished. 

\textbf{Differences from Other BERT-based Models.}
Most existing BERT-based models for clarifying question selection leverage the topic(query), questions, and answers in the conversation history and do not differentiate answers that are confirmation or denial \cite{aliannejadi2019asking, hashemi2020guided}. In contrast, MMR-BERT is specifically designed to leverage negative feedback from conversation history, which means it uses previously asked questions as input and does not use the answers in the history as they are all denial (we assume that the system would stop asking questions when it has identified the user intent). 
From the perspective of model design, existing models typically use average BERT representations of each historical conversation turn \cite{aliannejadi2019asking} or concatenate the sequence of a query, question, and answer in each turn as input to BERT models \cite{hashemi2020guided}. When used in the intent clarification task, these methods either do not differentiate the effect of each asked question or do not consider the effect of the initial query should be modeled differently from the questions with negative feedback. Following the MMR principle, our MMR-BERT model takes the task characteristics into account and thus can more effectively use negative feedback.
\section{Experimental Setup}
\label{sec:experiments}
This section introduces the data we use for experiments, how we evaluate the proposed models, the competing methods for comparison, and the technical details in the experiments. 

\subsection{Data}
\label{subsec:data}
We use Qulac \cite{aliannejadi2019asking} for experiments. As far as we know, it is the only dataset with mostly yes/no clarifying questions in information-seeking conversations. Qulac uses the topics in the TREC Web Track 2009-2012 \cite{clarke2009overview, clarke2012overview} as initial user queries. These topics are either ``ambiguous'' or ``faceted'' and are originally designed for the task of search result diversification. 
For each topic, Qulac has collected multiple clarifying questions for each facet (or intent) of the topic through crowd-sourcing; then for each facet of the topic, Qulac obtained the answers to all the questions of the topic from the annotators. The relevance judgments of documents regarding each topic-facet are inherited from the TREC Web track. 

We refined Qulac for the intent clarification task by assigning labels 2 or 1 to the questions that receive positive or negative feedback in the answers and label 0 to questions not associated with the topic. 
Many negative answers in Qulac also include the user's true intent, such as ``No. I want to know B.'' to the question ``Do you want to know A?''. 
It is too optimistic to assume users always provide true intents in their answers. Also, in that case, negative feedback does not have difference from positive feedback or is even better. 
To test how the models performs at incorporating negative feedback alone, we ignore the supplementary information and only keep ``no'' as user answers. 
For questions that are not yes/no questions, we consider the answers are negative feedback. 

To check whether a model can clarify user intents based on the negative feedback in the conversation history more sufficiently, we enlarge the dataset by including all the questions with label 1 as a 1-turn conversation for each topic-facet. In other words, besides letting the model select the first question, we also enumerate all the questions with label 1 as the first question to check how a model performs under various contexts. The original Qulac enumerates all the questions associated with a query to construct conversations of 1 to 3 turns and only select 1 more question based on the pre-constructed static conversation history. While we also enlarge the data similarly, we only construct conversations with 1 turn, and select questions based on previously selected questions. 

The resulting data has 8,962 conversations in total, including 762 conversations of 0-turn (only initial query) and 8,200 1-turn (the added conversations). With the enlarged data, we have many more conversations with various contexts as feedback to test the models and to establish the effectiveness of the results. The statistics are shown in Table \ref{tab:statistics}. 

\begin{table}
    \centering
    \small
    \caption{Statistics of our revised version of Qulac.}
    \scalebox{0.9}{
    \begin{tabular}{l l}
    \hline
    \# topics & 198 \\
    \# faceted/ambiguous topics & 141/57 \\
    \hline
    \# facets  &  762 \\
    Average/Median facet per topic  &  3.85$\pm$1.05/4 \\
    \# informational/navigational facets  & 577/185 \\
    \hline
    \# questions/question-answer pairs & 2,639/10,277 \\
    \# \textit{question with positive answers} & \textit{2,007} \\
    Average words per question/answer  & 9.49$\pm$2.53/8.21$\pm$4.42 \\
    \hline
    \# \textit{expanded conversations} & \textit{8,962} \\
    \# conversations starting with 0/1 turns & 762/8,200 \\
    \hline
    \end{tabular}
    }
    \label{tab:statistics}
\end{table}


\subsection{Evaluation}
\label{subsec:evaluation}
We evaluate the models on two tasks: 1) the proposed intent clarification task to see whether it can ask the questions covering the true user intent within fewer conversation turns; 2) the associated document retrieval task to see whether the asked clarifying questions can improve the document retrieval performance. Following \cite{aliannejadi2019asking, hashemi2020guided}, we use 5-fold cross-validation for evaluation. We split the topics to each fold according to their id modulo 5. Three folds are used for training, one fold for validation, and one fold for testing. For the question ranking task, we use Query Likelihood (QL) \cite{ponte1998language} to retrieve an initial set of candidates and conduct re-ranking with BERT-based models. For the document retrieval task, as in \cite{aliannejadi2019asking, hashemi2020guided}, we use the revised QL model for retrieval: replacing the original query language model with a convex combination of the language models of the initial query ($t$) and all the question-answer pairs in the conversation ($h$).

For the intent clarification task, we \textit{concatenate the question asked in each conversation turn as a ranking list} for evaluation. 
The primary evaluation metric is MRR calculated based on questions with label 2, which indicates \textbf{the number of turns} a model needs to identify true user intent. We also include NDCG@3 and NDCG@5 based on labels 2 and 0 to show how a model identifies the target questions in the first 3 or 5 interactions. 
To evaluate the overall quality of the clarifying questions, we also use NDCG@3 and NDCG@5 computed using the multi-grade labels 2, 1, and 0 as metrics. These metrics also give rewards to the questions that receive negative feedback from users but are still relevant to the topic.
We exclude NDCG@1 since the focus of the evaluation is to see how a model leverages the negative feedback in the context, whereas the first question is ranked based on only the original query. Also, the initial question in most of the conversations is with label 1 in the enlarged dataset regardless of the model used. 

For the document retrieval task, we use MRR, Precision(P)@1, NDCG@1, 5, and 20 as the evaluation metrics. 
MRR measures the position of the first relevant documents. NDCG@1, 5, and 20 indicate the performance based on 5-level labels (0-4) at different positions. 
Fisher random test \cite{smucker2007comparison} with $p<0.05$ is used to measure statistical significance for both tasks.

\subsection{Baselines}
\label{subsec:baseline}
We include seven representative baselines to select questions and compare their performance to MMR-BERT on both the intent clarification task and the associated document retrieval task: 

\textbf{QL}: The Query Likelihood \cite{ponte1998language} (QL) model is a term-based retrieval model that ranks candidates by the likelihood of a query generated from a candidate, also serving to collect initial candidates.

\textbf{BERT-INIT}: A BERT-based model trained with label 1 and 0 in Section \ref{subsec:init_retrieval}.

\textbf{MBERT-INIT}: A BERT-based model trained with label 2, 1 and 0 as mentioned in Section \ref{subsec:init_retrieval}.

\textbf{SingleNeg}\cite{karimzadehgan2011improving}: A negative feedback method that extracts a single negative topic model from the mixture with the language model of background corpus built with the non-relevant results.
\textbf{MMR}: The Maximal Marginal Relevance (MMR) model \cite{carbonell1998use} ranks questions according to the original MMR equation proposed for search diversification as \begin{equation}
\label{eq:mmr}
\arg max_{q\in Q \setminus Q_h} \lambda f(t,q) - (1-\lambda) max_{q'\in Q_h}f(q', q),
\end{equation}
where we set $f(.,.) = sigmoid(\text{BERT-INIT}(.,.))$ to measure similarity, and $0\leq\lambda\leq1$ is a hyper-parameter.


\textbf{BERT-NeuQS}: BERT-NeuQS \cite{ai2018learning} uses the \textit{average BERT representations of questions and answers in each historical conversation turn} as well as features from query performance prediction (QPP) for next clarifying question selection. To see the effect of model architecture alone, we did not include the QPP features. 

\textbf{BERT-GT}: The Guided Transformer model (BERT-GT) \cite{hashemi2020guided} encodes conversation history by inputting \textit{the concatenated sequence of a topic (query), clarifying questions and answers in the history} to a BERT model, guided by top-retrieved questions or documents to select next clarifying question.

QL, BERT-INIT, and MBERT-INIT only use the initial query for ranking while the other models also consider the conversation history. SingleNeg and MMR are based on heuristics. BERT-NeuQS and BERT-GT are state-of-the-art neural models for clarifying question selection. 
We discard the numbers of other negative feedback methods such as MultiNeg \cite{karimzadehgan2011improving} and Rocchio \cite{rocchio1971relevance} due to their inferior performance. 
BERT-NeuQS uses the query performance prediction scores of a candidate question for document retrieval to enrich the question representation. Our model significantly outperforms BERT-NeuQS if we also add this information. However, since we focus on studying which method is better at leveraging the negative feedback, for fair comparisons, we do not include this part for both BERT-NeuQS and our model. BERT-GT works better with questions than documents in our experiments so we only report the setting with questions. MMR-BERT uses the first question from BERT-INIT as its initial question. 

\subsection{Technical Details}
\label{subsec:tech}
We first fine-tuned the ``bert-base-uncased'' version of BERT \footnote{\url{https://github.com/huggingface/transformers}} using our local documents with 3 epochs. Then we fine-tuned BERT-INIT with 5 epochs allowing all the parameters to be updated. All the other BERT-based models loaded the parameters of the trained BERT-INIT and fixed the parameters in the transformer encoder layers during training. This is because the tremendous amount of parameters in the BERT encoders can easily overwhelm the remaining parameters in different models on the data at Qulac's scale, which makes the model performance unstable. The variance of model performance is huge in multiple runs if we let all the parameters free, which leads to unconvincing comparisons. 
The limit of conversation turns $k$ was set to 5. 
We optimized these models with the Adam \cite{kingma2014adam} optimizer and learning rate 0.0005 for 10 epochs. The number of MLP layers that have output dimension 1 was set from $\{1,2\}$. The dimension of the hidden layer of the 2-layer MLPs was selected from $\{4,8,16,32\}$. $\lambda$ in Equation \eqref{eq:mmr} and the query weight in SingleNeg were scanned from 0.8 to 0.99. Feedback term count in SingleNeg was chosen from $\{10,20,30\}$. Top 10 questions were used in BERT-GT. The coefficient to balance the weight of initial query and conversation history in the document retrieval model was scanned from 0 to 1 for each method.


\section{Results and Discussion}
\label{sec:results}
Next, we show the experimental results of the clarifying question selection task and the associated document retrieval task. We analyze the model behaviors as well as success and failure cases. 

\begin{table}
    \centering
    \caption{Model performance on intent clarification task evaluated using only label 2 or both label 1 \& 2. `*' indicates the best baseline results, and `$\dagger$' shows the statistically significant improvements over them. }
    \scalebox{0.95}{
    \begin{tabular}{l||l|l|l||l|l}
    \hline
    \multirow{2}{*}{Model} 
    & \multicolumn{3}{c||}{Label 2 only} & \multicolumn{2}{c}{Label 1\&2} \\
    \cline{2-6}
    & MRR & NDCG3 & NDCG5 & NDCG3 & NDCG5 \\
    \hline
    QL & 0.216 & 0.130 & 0.159 & 0.514 & 0.565 \\
    BERT-INIT & 0.235 & 0.143 & 0.173 & 0.531 & 0.583 \\
    MBERT-INIT & 0.235 & 0.144 & 0.173 & 0.532* & 0.584 \\
    \hline
    SingleNeg & 0.217 & 0.131 & 0.160 & 0.513 & 0.565 \\
    MMR & 0.237 & 0.144 & 0.178 & 0.531 & 0.585* \\
    BERT-NeuQS & 0.241 & 0.146 & 0.182* & 0.528 & 0.580 \\
    BERT-GT & 0.242* & 0.148* & 0.178 & 0.530 & 0.580 \\
    \hline
    MMR-BERT & \textbf{0.248$^{\dagger}$} & \textbf{0.152$^{\dagger}$} & \textbf{0.189$^{\dagger}$} & \textbf{0.533} & \textbf{0.586$^{\dagger}$} \\
    \hline
    \end{tabular}
    }
    \label{tab:cq_results}
\end{table}

\subsection{Clarifying Question Selection Results}
\label{subsec:cq_results}
\textbf{Overall Performance.}
As shown in Table \ref{tab:cq_results}, MMR-BERT has achieved the best performance to identify the target questions that cover true user intents. It outperforms the best baselines significantly regarding almost all the metrics. 
Note that the evaluation is based on 8,962 conversations and 8,200 of them have the same first negative question in the enlarged data so all the models can refine the question selection only from the second question for most conversations. This limits the improvements of MMR-BERT over the baselines. However, the improvements on about nine thousand data points are significant.

Word-based methods (QL and SingleNeg) are inferior to the other neural methods by a large margin. Also, SingleNeg hardly improves upon QL, indicating that word-based topic modeling methods are not effective to incorporate negative feedback in clarifying question selection, probably due to insufficient words to build topic models. The BERT-based methods using the feedback information can identify the first target questions earlier than BERT-INIT and MBERT-INIT.
With the similarity function provided by BERT-INIT, MMR can outperform BERT-INIT. The ability of BERT models to measure semantic similarity is essential for the MMR principle to be effective. 
Moreover, while BERT-NeuQS and BERT-GT improve the metrics regarding label 2, their performance regarding questions with label 1 is harmed. BERT-NeuQS concatenates the topic representation with the average representations of each q-a pair and BERT-GT encode the sequence of the conversation history ($t,(q_1,a_1),\cdots, (q_{|h|},a_{|h|})$) as a whole. Thus it could be difficult for them to figure out which part a candidate question should be similar to and which part not. By matching a candidate question with the topic and each historical question individually, MMR-BERT can balance the similarity to the topic and dissimilarity to the historical questions better. 

\begin{figure}
	\includegraphics[width=0.4\textwidth]{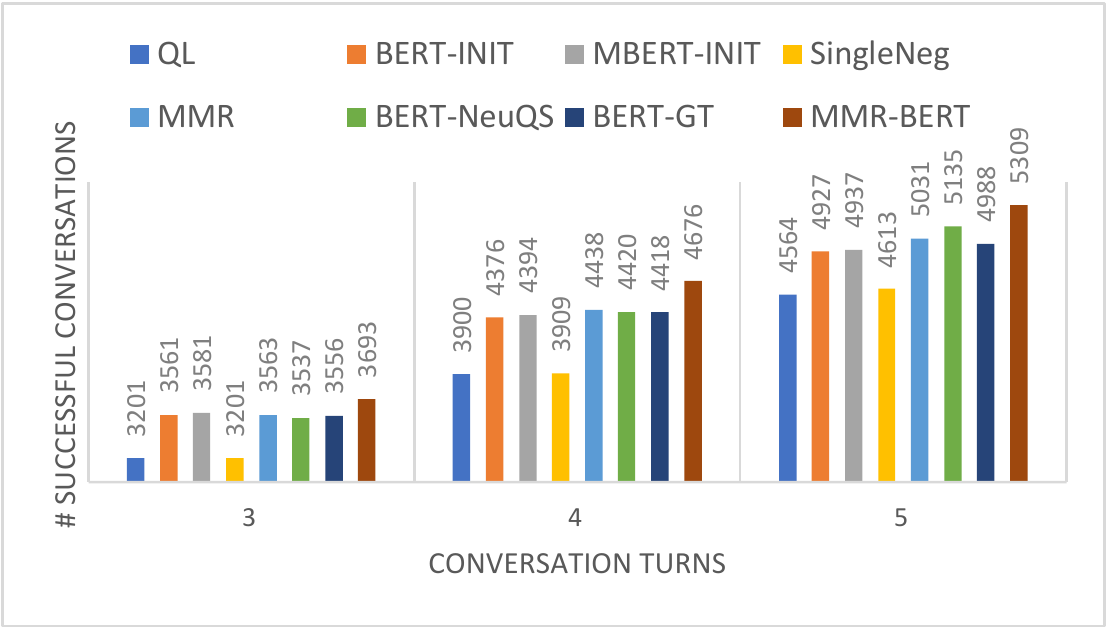} %
	\caption{Comparison of MMR-BERT and baselines in terms of the cumulative number of success conversations at each turn on the intent clarification task.}
	\label{fig:succ_conv_turns}
\end{figure}

\textbf{Number of Success Conversations.}
Figure \ref{fig:succ_conv_turns} shows the cumulative number of success conversations of each method that correctly identifies user intents at the third, fourth, and fifth turns. 
We focus more on how to leverage the negative feedback in the conversation so far rather than how to ask the first clarifying question without feedback information. As shown in the figure, among all the 8,962 conversations, MMR-BERT identifies user intents in 41.2\%, 52.2\%, and 59.2\% conversations by asking at most 3, 4, and 5 clarifying questions. The best baseline at each turn is different while MMR-BERT always has the overall best performance across various turns. This indicates that our MMR-BERT can leverage negative feedback more effectively than the baselines in identifying user intents. 

\begin{figure}
	\includegraphics[width=0.35\textwidth]{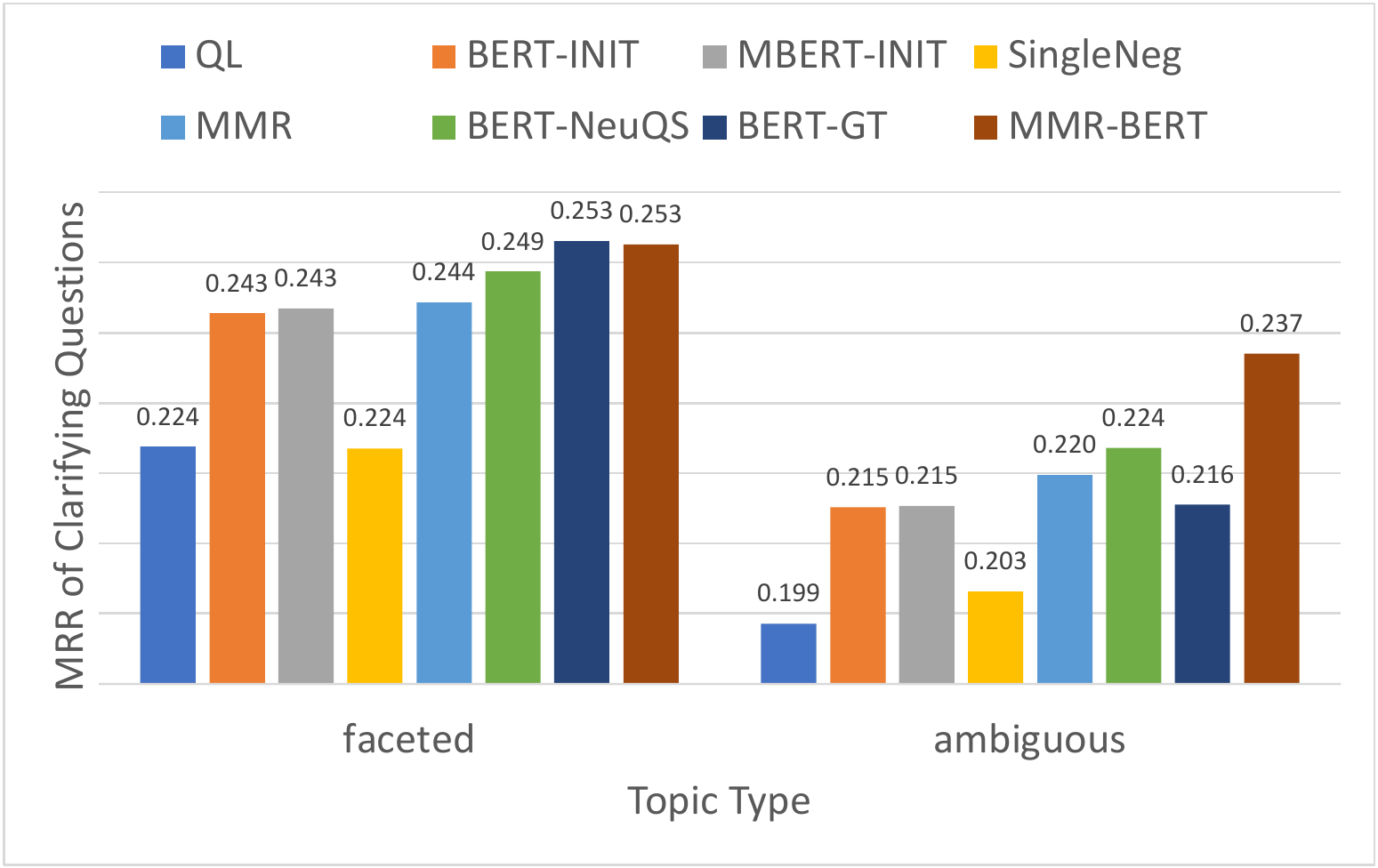} %
	\caption{MRR of each method in the intent clarification task in terms of topic type. }
	\label{fig:mrr_topic_type}
\end{figure}

\textbf{Impact of Topic Type.}
In Figure \ref{fig:mrr_topic_type}, we study how MMR-BERT performs on queries of different types compared with other methods. As we mentioned in Section \ref{subsec:data}, query topics in Qulac are faceted or ambiguous. An example of a faceted query is ``elliptical trainer'', which has the facets such as ``What are the benefits of an elliptical trainer compared to other fitness machines?'', ``where can I buy a used or discounted elliptical trainer?'', ``What are the best elliptical trainers for home use?'' and ``I'm looking for reviews of elliptical machines.'' An ambiguous query is a query that has multiple meanings, e.g., ``memory'', which can refer to human memory, computer memory, and the board game named as memory. From Figure \ref{fig:mrr_topic_type}, we have two major observations:

1) All the methods perform better on faceted queries than on ambiguous queries. 
Since QL performs worse on ambiguous queries than on faceted queries by a large margin, the performance of other methods is limited by the quality of initial candidate clarifying questions retrieved by QL. It also indicates that questions for ambiguous queries in the corpus have less word matching than faceted queries.

2) The improvements of MMR-BERT over other methods are much larger on ambiguous queries than on faceted queries. It is essential to differentiate the semantic meanings of various clarifying questions relevant to the same query when leveraging the negative feedback. Clarifying questions of a faceted query are usually about subtopics under the small space of the query topic and the words co-occurring with the query in each subtopic have much overlap. Again for the ``elliptical trainer'' example, the latter associated 3 intents are all related to the purchase need, and the words such as ``buy'', ``best'', and ``reviews'' can co-occur often in the corpus. Thus it is difficult to differentiate these questions even for BERT-based models. In contrast, clarifying questions corresponding to each meaning of an ambiguous query usually consist of different sets of context words, e.g., human memory can have ``memory loss'' and ``brain'' in the related texts while computer memory always co-occurs with ``disk'', ``motherboard'', etc. 
As BERT has seen various contexts in a huge corpus during pre-training, they have better capabilities to differentiate the meanings of an ambiguous query compared to the subtopics of a faceted query. However, BERT-NeuQS and BERT-GT cannot fully take advantage of BERT's ability to differentiate semantic meanings due to their architecture, either averaging the representations of historical questions or encoding the sequence of query and the asked questions. 


\begin{figure}
	\includegraphics[width=0.35\textwidth]{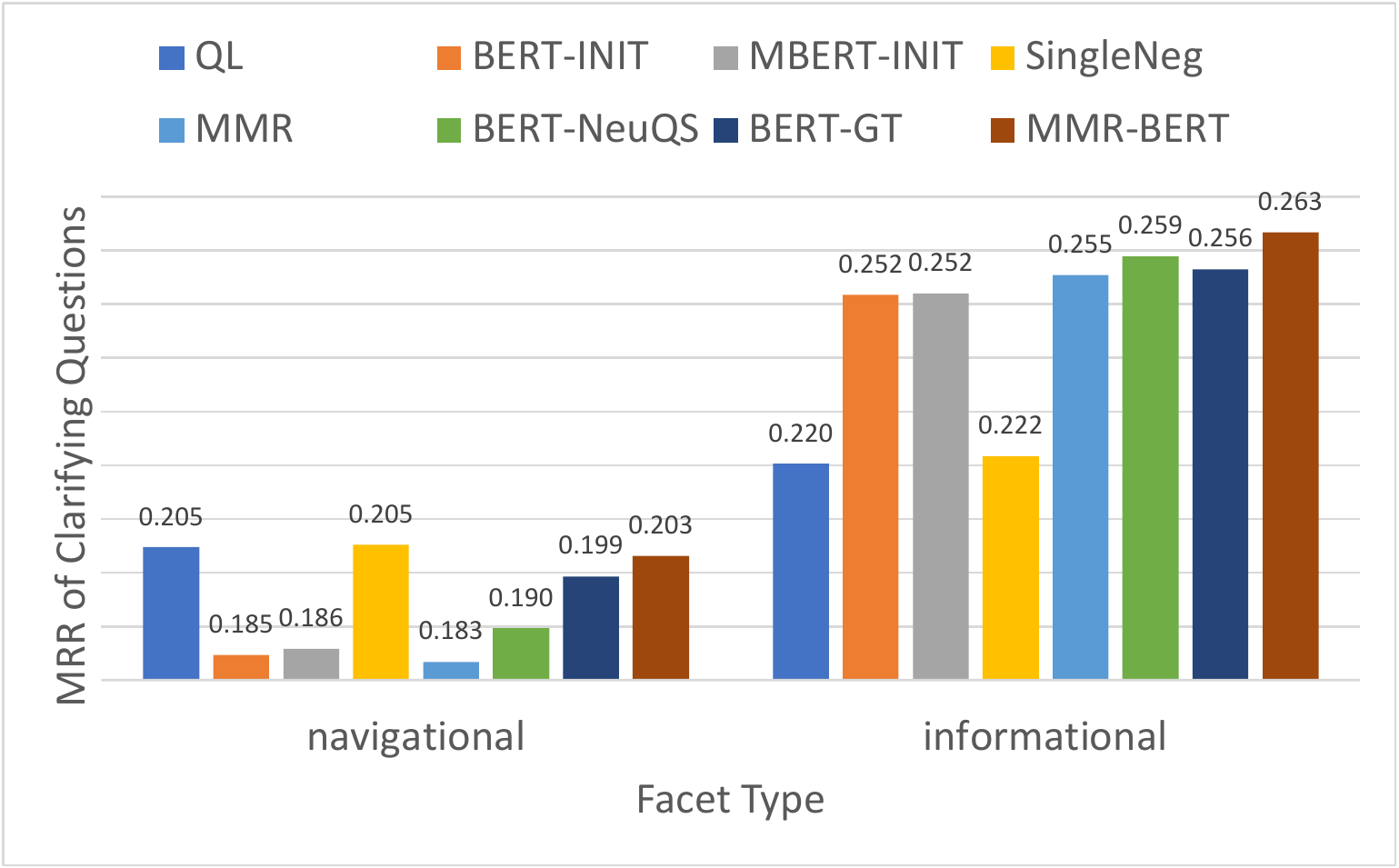} %
	\caption{MRR of each method in the intent clarification task in terms of facet type. }
	\label{fig:mrr_facet_type}
\end{figure}

\textbf{Impact of Facet Type.}
We compare each method in terms of their performance on different types of intent facets in Figure \ref{fig:mrr_facet_type}. Similar to the varied performance in terms of topic type, QL performs worse on navigational facets than on informational facets. The clarifying questions that ask about navigational intents sometimes do not match any of the query words such as ``are you looking for a specific web site?'' and ``any specific company on your mind?'' In such cases, the target questions are not included in the candidate pool for re-ranking, which leads to inferior performance on navigational queries. 

In addition, we find that neural methods perform worse than word-matching-based methods on navigational queries. Questions that ask about navigational intents are usually in the format of ``do you need any specific web page about X (query)?'' rather than the typical format of questions about informational intents such as ``are you interested in Y (subtopics) of X (query)?'' Also, navigational facets are much fewer than informational facets (185 versus 577), which leads to a smaller amount of questions about navigational facets. The supervised neural models tend to promote questions asking about informational intents during re-ranking since they are semantically more similar to the query (talking about their subtopics) and they are more likely to be relevant in the training data. In contrast, word-matching-based methods treat navigational and informational questions similarly since they both hit query words and have similar length. By selecting the next question different from previous questions and relevant to the query, MMR-BERT does not demote questions about navigational facets and does not harm the performance on navigational facets.

\subsection{Document Retrieval Performance}
\label{subsec:conv_doc_perf}

\begin{table}
    \centering
    \caption{Document retrieval performance with conversations composed by each model. The best baseline results are marked with `*', and the statistically significant improvements over them are marked with`$\dagger$'.}
    \scalebox{0.96}{
    \begin{tabular}{l|l|l|l|l|l}
    \hline
    Model & MMR & P1 & NDCG1 & NDCG5 & NDCG20 \\
    \hline
    OriginalQuery & 0.267 & 0.181 & 0.121 & 0.128 & 0.131 \\
    \hline
    QL & 0.292 & 0.209 & 0.146 & 0.142 & 0.141 \\
    BERT-INIT & 0.299 & 0.210* & 0.145 & 0.143 & 0.143 \\
    MBERT-INIT & 0.298 & 0.209 & 0.143 & 0.142 & 0.144* \\
    \hline
    SingleNeg & 0.292 & 0.209 & 0.147* & 0.142 & 0.141 \\
    MMR & 0.301* & 0.210* & 0.143 & 0.143 & 0.144* \\
    BERT-NeuQS & 0.296 & 0.209 & 0.145 & 0.145* & 0.142 \\
    BERT-GT & 0.294 & 0.206 & 0.141 & 0.145* & 0.143 \\
    \hline
    MMR-BERT & \textbf{0.306$^{\dagger}$} & \textbf{0.217$^{\dagger}$} & \textbf{0.151$^{\dagger}$} & \textbf{0.146} & \textbf{0.146$^{\dagger}$} \\
    \hline
    \end{tabular}
    }
    \label{tab:doc_results}
\end{table}

\begin{figure}
	\includegraphics[width=0.35\textwidth]{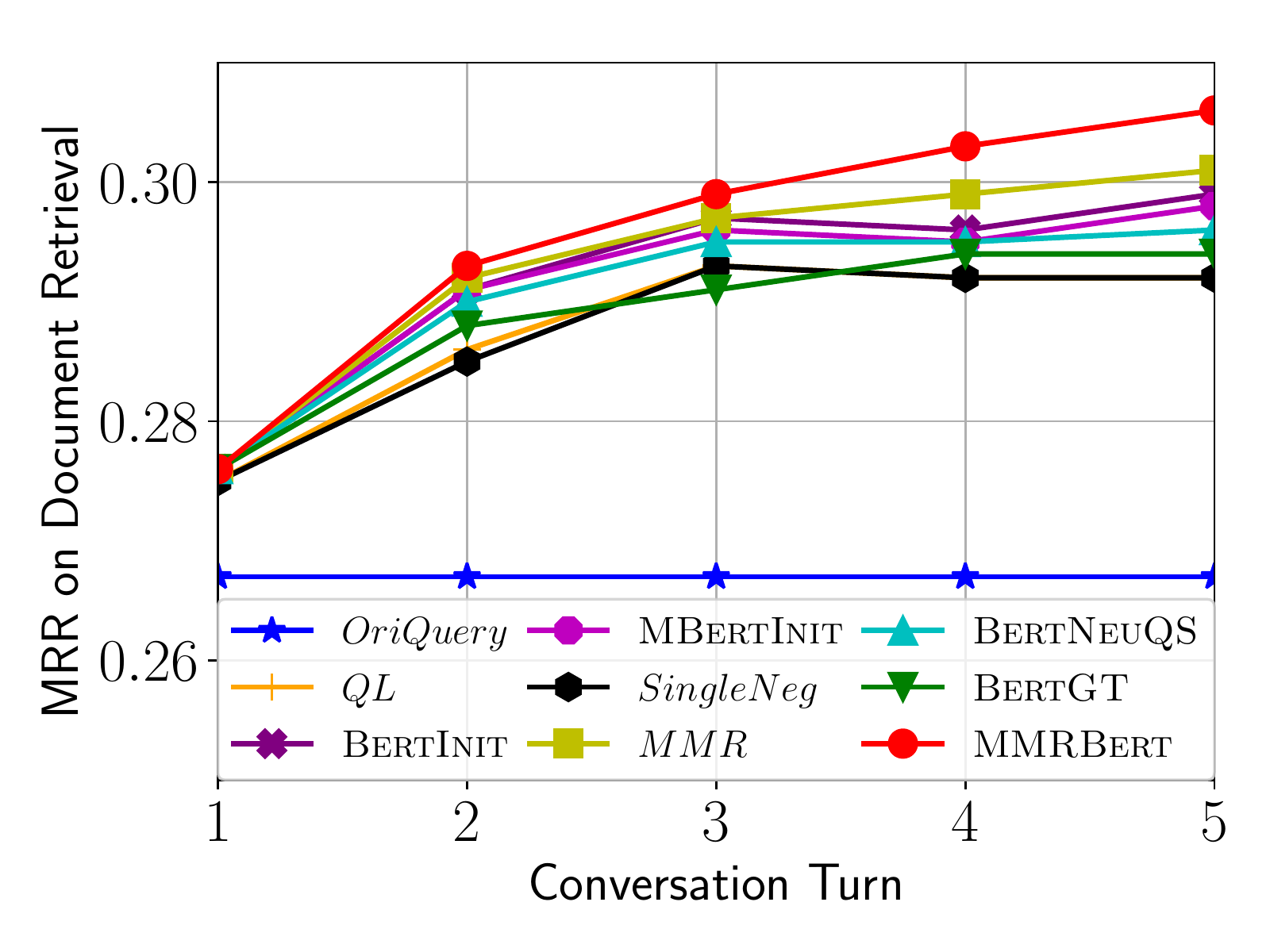} %
	\caption{MRR at each turn on document retrieval. }
	\label{fig:doc_turns_mrr}
\end{figure}

\label{subsec:doc_results}
Table \ref{tab:doc_results} and Figure \ref{fig:doc_turns_mrr} show the document retrieval performance of using the original query alone and using the conversations produced by each method. In Table \ref{tab:doc_results}, we observe that all the question selection methods can promote relevant documents significantly by asking clarifying questions. The questions asked by MMR-BERT achieve the best document retrieval performance, indicating that our model can find users' target information at higher positions by identifying user intents better. 
Since the model for document retrieval is a simple word-based model, the advantage of asking correct questions may not be reflected in retrieving documents. The cases in Section \ref{subsec:cq_doc_mrr_case} show this point. Also, as mentioned in Section \ref{subsec:tech}, the methods can ask at most 5 questions when they cannot identify user intents. These questions could have more supplementary information than BERT-MMR in finding relevant documents if they are of label 1. Nonetheless, 
MMR-BERT still achieves significant improvements on 8,962 conversations. 

Figure \ref{fig:doc_turns_mrr} confirms the advantage of MMR-BERT by showing that it can retrieve documents relevant to user needs better at earlier turns as well. With more interactions allowed, MMR-BERT can identify more true user intents and thus achieve better document retrieval performance. 
Among the baselines that select questions using negative feedback, MMR has the best evaluation results most of the time, probably due to its better overall performance in intent clarification, shown in Table \ref{tab:cq_results}. It boosts questions with label 2 without harming the performance of questions with label 1. Using revised QL for document retrieval, questions of label 1 can also be more helpful than a non-relevant question.

\begin{table*}
    \centering
    \footnotesize
    \caption{Good and bad cases of MMR-BERT compared with the best baseline - BERT-GT in terms of their MRR differences($\Delta$MRR of CQ) in the intent clarification task. The maximal number of conversation turns is 5. $\Delta$MRR of Doc denotes the MRR difference of the associated document retrieval task after the conversation. Queries are shown in the format of \textit{query(facet description); topic type; facet type}.   } %
    \scalebox{0.935}{
    \begin{tabular}{l|l|p{2.2cm}}
    \hline
    \multicolumn{3}{c}{Query: ``\textbf{diversity}''(``\textbf{How is workplace diversity achieved and managed}?''); \textbf{faceted}; \textbf{informational}} \\
    \hline
    \multirow{5}{*}{BERT-GT}
    & are you looking for a definition of diversity? no & \multirow{7}{*}{\parbox{2.2cm}{$\Delta$MRR of CQ: +0.500 $\Delta$MRR of Doc: +0.667}} \\
    & would you like the legal definition of diversity? no & \\
    & would you like to know how diversity helps or harms an organization? no & \\
    & do you need the definition of diversity? no & \\
    & would you like to see results about diversity in a business setting? no & \\
    \cline{1-2}
    \multirow{2}{*}{MMR-BERT}
    & are you looking for a definition of diversity? no \\
    & are you looking for educational materials about diversity? yes, i need materials on achieving workplace diversity \\
    \hline
    \multicolumn{3}{c}{Query: \textbf{``flushing''}(``\textbf{Find information about Flushing, a neighborhood in New York City.}''); \textbf{ambiguous}; \textbf{informational} } \\
    \hline
    \multirow{5}{*}{BERT-GT}
    & would you like to see diagrams of how a toilet flushes? no & \multirow{7}{*}{\parbox{2.2cm}{$\Delta$MRR of CQ: +0.500 $\Delta$MRR of Doc: +0.005}}\\
    & would you like to know about the plumbing mechanisms of flushing? no \\
    & what aspect of the flushing remonstrance would you like to learn more about? no \\
    & which battle would you like to learn about how the technique of flushing was used? no \\
    & what flushing are you interested in toilet or facial? no \\
    \cline{1-2}
    \multirow{2}{*}{MMR-BERT}
    & would you like to see diagrams of how a toilet flushes? no \\
    & are you referring to flushing new york? yes \\
    \hline
    \multicolumn{3}{c}{Query: ``\textbf{the sun}''(``\textbf{Find the homepage for the U.K. newspaper, The Sun.}''); \textbf{ambiguous}; \textbf{navigational}} \\
    \hline
    \multirow{5}{*}{BERT-GT}
    & are you interested in the suns size? no & \multirow{7}{*}{\parbox{2.2cm}{$\Delta$MRR of CQ: +0.500 $\Delta$MRR of Doc: +0.000}} \\
    & are you interested in objects orbiting the sun? no \\
    & do you want to know how far the sun is away from different planets? no \\
    & are you looking for information about how big the sun is? no \\
    & do you want to know facts about the sun? no \\
    \cline{1-2}
    \multirow{2}{*}{MMR-BERT}
    & are you interested in the suns size? no \\
    & are you looking for the news paper the sun? yes, the uk newspaper \\
    \hline
    \hline
    \multicolumn{3}{c}{Query: ``\textbf{raised gardens}''(``\textbf{Find photos of raised garden beds.}''); \textbf{faceted}; \textbf{navigational}} \\
    \hline
    \multirow{2}{*}{BERT-GT}
    & do you need information in different types that can be made? no & \multirow{7}{*}{\parbox{2.2cm}{$\Delta$MRR of CQ: -0.500 $\Delta$MRR of Doc: -0.166}}\\
    & is your request related to raised garden beds? yes, find pictures of it \\
    \cline{1-2}
    \multirow{5}{*}{MMR-BERT}
    & do you need information in different types that can be made? no \\
    & what specific supply would you like to buy for your raised garden? no \\
    & do you want to take a class about raised gardens? no \\
    & do you want to buy a book about raised? no \\
    & do you want to know how to create a raised garden? no \\
    \hline
    \multicolumn{3}{c}{Query: ``\textbf{rice}''(``\textbf{Find recipes for rice, for example fried rice or rice pudding.}''); \textbf{ambiguous}; \textbf{informational} } \\
    \hline
    \multirow{2}{*}{BERT-GT}
    & are you looking for a specific type of rice? no  & \multirow{7}{*}{\parbox{2.2cm}{$\Delta$MRR of CQ: -0.500 $\Delta$MRR of Doc: -0.000}}\\
    & are you looking for recipes that include rice? yes, i want recipes for rice \\
    \cline{1-2}
    \multirow{5}{*}{MMR-BERT}
    & are you looking for a specific type of rice? no \\
    & are you looking for rice university? no \\
    & do you want to know the nutritional content of rice? no \\
    & are you referring to a person named rice? no \\
    & what type of rice dish are you looking? no \\ 
    \hline
    \multicolumn{3}{c}{Query: ``\textbf{flushing}''(``\textbf{Find a street map of Flushing, NY.}''); \textbf{ambiguous}; \textbf{navigational}} \\
    \hline
    \multirow{2}{*}{BERT-GT}
    & would you like directions to flushing new york? no  & \multirow{7}{*}{\parbox{2.2cm}{$\Delta$MRR of CQ: -0.500 $\Delta$MRR of Doc: -0.167}}\\
    & are you referring to flushing new york? yes, exactly \\
    \cline{1-2}
    \multirow{5}{*}{MMR-BERT}
    & would you like directions to flushing new york? no \\
    & would you like to know about the plumbing mechanisms of flushing? no \\
    & do you want to know why your face is flushing? no \\
    & are you looking for a directions to the new york hall of science in flushing meadows corona park? no \\
    & which battle would you like to learn about how the technique of flushing was used? no \\
    \hline
    \end{tabular}
    }
    \label{tab:mmr_bert_cases}
\end{table*}

\subsection{Case Analysis}
\label{subsec:cq_doc_mrr_case}
We extract some representative successful and failure cases of MMR-BERT compared with the best baseline - BERT-GT in terms of MRR in the intent clarification task, shown in Table \ref{tab:mmr_bert_cases}. We include conversations of faceted and ambiguous queries as well as navigational and informational facets for both good and bad cases to show how the models perform on various types of queries and facets. In these cases, MMR-BERT and BERT-GT have the same initial clarifying questions with negative feedback. These cases show how MMR-BERT and BERT-GT select the next question based on the same previous negative feedback. 

\textbf{Success Cases.}
MMR-BERT identifies the correct user intent by selecting questions that are relevant to the query while different from previous questions with negative feedback. In contrast, BERT-GT tends to select questions that are similar to both the query and the previous questions. For the example query ``diversity'', the initial clarifying question asks whether the intent is to find the definition of diversity. MRR-BERT asks the user whether he/she needs the educational materials about diversity in the second turn. However, BERT-GT still asks questions about the definition of diversity twice in the following four turns. For the ambiguous query ``flushing'', given negative feedback on the first question about toilet flushing, MMR-BERT asks about Flushing in New York in the next question while BERT-GT still asks about the flushing of the same meaning in the second question. For another ambiguous query ``the sun'', the first clarifying question is about sun size. Based on the negative response, MMR-BERT asks about another meaning of the sun - the newspaper named as the sun. In contrast, the next four questions BERT-GT asks are all about the sun as a star, and the question in the fourth turn is again about the size of the sun. Improvements in identifying the correct clarifying questions can lead to better performance in the associated document retrieval task but it is not always the case probably due to the simplicity of the document retrieval model.  

\textbf{Failure Cases.}
The questions asked by MMR-BERT in each conversation are more diverse and tend to cover more intents. However, the questions that receive positive feedback sometimes are more semantically similar to the questions with negative feedback than the other questions. In such cases, MMR-BERT fails to identify the correct intents within fewer conversation turns by asking diverse questions. For the faceted query ``raised gardens'' with intent ``find photos of raised garden beds'', the initial question does not include any query words, so emphasizing the difference from this question is not helpful and could even be harmful to select next question by introducing noise. For the ambiguous query ``rice'', the first question asking whether the user wants a specific type of rice receives a negative response. In the following conversations, MMR-BERT asks about other meanings of rice such as Rice University and a person named Rice. BERT-GT selects the question that is also related to the meaning of rice as food in the next turn. Although referring to the same meaning, the aspect of the recipe is the true user intent. Similarly, for the query ``flushing'', while the user wants the street map of Flushing New York, the question that asks about the direction to Flushing New York receives negative feedback. MMR-BERT selects questions about other meanings of flushing in the next several turns including the mechanism or technique, face flushing, and Flushing meadows corona park. However, the true intent is another facet of the same meaning. These cases argue for other strategies to ask questions such as clarifying meanings for ambiguous queries first and then asking about the subtopics under the correct meaning. We leave this study as future work. The performance of MMR-BERT in these cases in the associated document retrieval task sometimes is not always worse than BERT-GT, due to some useful information contained in the conversations even though the questions do not receive positive feedback. 


\section{Conclusion}
In this paper, we propose an intent clarification task based on yes/no clarifying questions in information-seeking conversations. The task's goal is to ask questions that can uncover the true user intent behind an ambiguous or faced query within the fewest conversation turns. 
We propose a maximal-marginal-relevance-based BERT model (MMR-BERT) that leverages the negative feedback to the previous questions using the MMR principle. 
Experimental results on the refined Qulac dataset show that MMR-BERT has significantly better performance than the competing question selection models in both the intent identification task and the associated document retrieval task. 

For future work, we plan to evaluate the effect of the asked clarifying questions on the associated document retrieval task with neural document retrieval models. We are also interested in studying how to effectively use negative feedback on the clarifying questions in the document retrieval model. 

\begin{acks}
This work was supported in part by the Center for Intelligent Information Retrieval and in part by NSF IIS-1715095. Any opinions, findings and conclusions or recommendations expressed in this material are those of the authors and do not necessarily reflect those of the sponsor.
\end{acks}

\bibliographystyle{ACM-Reference-Format}
\balance
\bibliography{reference}


\begin{thebibliography}{40}


\ifx \showCODEN    \undefined \def \showCODEN     #1{\unskip}     \fi
\ifx \showDOI      \undefined \def \showDOI       #1{#1}\fi
\ifx \showISBNx    \undefined \def \showISBNx     #1{\unskip}     \fi
\ifx \showISBNxiii \undefined \def \showISBNxiii  #1{\unskip}     \fi
\ifx \showISSN     \undefined \def \showISSN      #1{\unskip}     \fi
\ifx \showLCCN     \undefined \def \showLCCN      #1{\unskip}     \fi
\ifx \shownote     \undefined \def \shownote      #1{#1}          \fi
\ifx \showarticletitle \undefined \def \showarticletitle #1{#1}   \fi
\ifx \showURL      \undefined \def \showURL       {\relax}        \fi
\providecommand\bibfield[2]{#2}
\providecommand\bibinfo[2]{#2}
\providecommand\natexlab[1]{#1}
\providecommand\showeprint[2][]{arXiv:#2}

\bibitem[\protect\citeauthoryear{Ai, Bi, Guo, and Croft}{Ai
  et~al\mbox{.}}{2018}]%
        {ai2018learning}
\bibfield{author}{\bibinfo{person}{Qingyao Ai}, \bibinfo{person}{Keping Bi},
  \bibinfo{person}{Jiafeng Guo}, {and} \bibinfo{person}{W~Bruce Croft}.}
  \bibinfo{year}{2018}\natexlab{}.
\newblock \showarticletitle{Learning a Deep Listwise Context Model for Ranking
  Refinement}.
\newblock \bibinfo{journal}{\emph{arXiv preprint arXiv:1804.05936}}
  (\bibinfo{year}{2018}), \bibinfo{pages}{135--144}.
\newblock


\bibitem[\protect\citeauthoryear{Aliannejadi, Kiseleva, Chuklin, Dalton, and
  Burtsev}{Aliannejadi et~al\mbox{.}}{2020}]%
        {aliannejadi2020convai3}
\bibfield{author}{\bibinfo{person}{Mohammad Aliannejadi},
  \bibinfo{person}{Julia Kiseleva}, \bibinfo{person}{Aleksandr Chuklin},
  \bibinfo{person}{Jeff Dalton}, {and} \bibinfo{person}{Mikhail Burtsev}.}
  \bibinfo{year}{2020}\natexlab{}.
\newblock \showarticletitle{ConvAI3: Generating Clarifying Questions for
  Open-Domain Dialogue Systems (ClariQ)}.
\newblock \bibinfo{journal}{\emph{arXiv preprint arXiv:2009.11352}}
  (\bibinfo{year}{2020}).
\newblock


\bibitem[\protect\citeauthoryear{Aliannejadi, Zamani, Crestani, and
  Croft}{Aliannejadi et~al\mbox{.}}{2019}]%
        {aliannejadi2019asking}
\bibfield{author}{\bibinfo{person}{Mohammad Aliannejadi},
  \bibinfo{person}{Hamed Zamani}, \bibinfo{person}{Fabio Crestani}, {and}
  \bibinfo{person}{W~Bruce Croft}.} \bibinfo{year}{2019}\natexlab{}.
\newblock \showarticletitle{Asking clarifying questions in open-domain
  information-seeking conversations}. In \bibinfo{booktitle}{\emph{Proceedings
  of the 42nd international acm sigir conference on research and development in
  information retrieval}}. \bibinfo{pages}{475--484}.
\newblock


\bibitem[\protect\citeauthoryear{Allan}{Allan}{2005}]%
        {allan2005hard}
\bibfield{author}{\bibinfo{person}{James Allan}.}
  \bibinfo{year}{2005}\natexlab{}.
\newblock \bibinfo{booktitle}{\emph{HARD track overview in TREC 2003 high
  accuracy retrieval from documents}}.
\newblock \bibinfo{type}{{T}echnical {R}eport}.
  \bibinfo{institution}{MASSACHUSETTS UNIV AMHERST CENTER FOR INTELLIGENT
  INFORMATION RETRIEVAL}.
\newblock


\bibitem[\protect\citeauthoryear{Belkin, Cool, Stein, and Thiel}{Belkin
  et~al\mbox{.}}{1995}]%
        {belkin1995cases}
\bibfield{author}{\bibinfo{person}{Nicholas~J Belkin}, \bibinfo{person}{Colleen
  Cool}, \bibinfo{person}{Adelheit Stein}, {and} \bibinfo{person}{Ulrich
  Thiel}.} \bibinfo{year}{1995}\natexlab{}.
\newblock \showarticletitle{Cases, scripts, and information-seeking strategies:
  On the design of interactive information retrieval systems}.
\newblock \bibinfo{journal}{\emph{Expert systems with applications}}
  \bibinfo{volume}{9}, \bibinfo{number}{3} (\bibinfo{year}{1995}),
  \bibinfo{pages}{379--395}.
\newblock


\bibitem[\protect\citeauthoryear{Bi, Ai, Zhang, and Croft}{Bi
  et~al\mbox{.}}{2019}]%
        {bi2019conversational}
\bibfield{author}{\bibinfo{person}{Keping Bi}, \bibinfo{person}{Qingyao Ai},
  \bibinfo{person}{Yongfeng Zhang}, {and} \bibinfo{person}{W~Bruce Croft}.}
  \bibinfo{year}{2019}\natexlab{}.
\newblock \showarticletitle{Conversational product search based on negative
  feedback}. In \bibinfo{booktitle}{\emph{CIKM'19}}. \bibinfo{pages}{359--368}.
\newblock


\bibitem[\protect\citeauthoryear{Carbonell and Goldstein}{Carbonell and
  Goldstein}{1998}]%
        {carbonell1998use}
\bibfield{author}{\bibinfo{person}{Jaime Carbonell} {and} \bibinfo{person}{Jade
  Goldstein}.} \bibinfo{year}{1998}\natexlab{}.
\newblock \showarticletitle{The use of MMR, diversity-based reranking for
  reordering documents and producing summaries}. In
  \bibinfo{booktitle}{\emph{Proceedings of the 21st annual international ACM
  SIGIR conference on Research and development in information retrieval}}.
  \bibinfo{pages}{335--336}.
\newblock


\bibitem[\protect\citeauthoryear{Cho, Zhang, Rao, Brockett, and Lee}{Cho
  et~al\mbox{.}}{2019}]%
        {cho2019generating}
\bibfield{author}{\bibinfo{person}{Woon~Sang Cho}, \bibinfo{person}{Yizhe
  Zhang}, \bibinfo{person}{Sudha Rao}, \bibinfo{person}{Chris Brockett}, {and}
  \bibinfo{person}{Sungjin Lee}.} \bibinfo{year}{2019}\natexlab{}.
\newblock \showarticletitle{Generating a Common Question from Multiple
  Documents using Multi-source Encoder-Decoder Models}.
\newblock \bibinfo{journal}{\emph{arXiv preprint arXiv:1910.11483}}
  (\bibinfo{year}{2019}).
\newblock


\bibitem[\protect\citeauthoryear{Choi, He, Iyyer, Yatskar, Yih, Choi, Liang,
  and Zettlemoyer}{Choi et~al\mbox{.}}{2018}]%
        {choi2018quac}
\bibfield{author}{\bibinfo{person}{Eunsol Choi}, \bibinfo{person}{He He},
  \bibinfo{person}{Mohit Iyyer}, \bibinfo{person}{Mark Yatskar},
  \bibinfo{person}{Wen-tau Yih}, \bibinfo{person}{Yejin Choi},
  \bibinfo{person}{Percy Liang}, {and} \bibinfo{person}{Luke Zettlemoyer}.}
  \bibinfo{year}{2018}\natexlab{}.
\newblock \showarticletitle{Quac: Question answering in context}.
\newblock \bibinfo{journal}{\emph{arXiv preprint arXiv:1808.07036}}
  (\bibinfo{year}{2018}).
\newblock


\bibitem[\protect\citeauthoryear{Clarke, Craswell, and Soboroff}{Clarke
  et~al\mbox{.}}{2009}]%
        {clarke2009overview}
\bibfield{author}{\bibinfo{person}{Charles~L Clarke}, \bibinfo{person}{Nick
  Craswell}, {and} \bibinfo{person}{Ian Soboroff}.}
  \bibinfo{year}{2009}\natexlab{}.
\newblock \bibinfo{booktitle}{\emph{Overview of the trec 2009 web track}}.
\newblock \bibinfo{type}{{T}echnical {R}eport}. \bibinfo{institution}{WATERLOO
  UNIV (ONTARIO)}.
\newblock


\bibitem[\protect\citeauthoryear{Clarke, Craswell, and Voorhees}{Clarke
  et~al\mbox{.}}{2012}]%
        {clarke2012overview}
\bibfield{author}{\bibinfo{person}{Charles~L Clarke}, \bibinfo{person}{Nick
  Craswell}, {and} \bibinfo{person}{Ellen~M Voorhees}.}
  \bibinfo{year}{2012}\natexlab{}.
\newblock \bibinfo{booktitle}{\emph{Overview of the TREC 2012 web track}}.
\newblock \bibinfo{type}{{T}echnical {R}eport}. \bibinfo{institution}{NATIONAL
  INST OF STANDARDS AND TECHNOLOGY GAITHERSBURG MD}.
\newblock


\bibitem[\protect\citeauthoryear{Croft and Thompson}{Croft and
  Thompson}{1987}]%
        {croft1987i3r}
\bibfield{author}{\bibinfo{person}{W~Bruce Croft} {and}
  \bibinfo{person}{Roger~H Thompson}.} \bibinfo{year}{1987}\natexlab{}.
\newblock \showarticletitle{I3R: A new approach to the design of document
  retrieval systems}.
\newblock \bibinfo{journal}{\emph{Journal of the american society for
  information science}} \bibinfo{volume}{38}, \bibinfo{number}{6}
  (\bibinfo{year}{1987}), \bibinfo{pages}{389--404}.
\newblock


\bibitem[\protect\citeauthoryear{Devlin, Chang, Lee, and Toutanova}{Devlin
  et~al\mbox{.}}{2018}]%
        {devlin2018bert}
\bibfield{author}{\bibinfo{person}{Jacob Devlin}, \bibinfo{person}{Ming-Wei
  Chang}, \bibinfo{person}{Kenton Lee}, {and} \bibinfo{person}{Kristina
  Toutanova}.} \bibinfo{year}{2018}\natexlab{}.
\newblock \showarticletitle{Bert: Pre-training of deep bidirectional
  transformers for language understanding}.
\newblock \bibinfo{journal}{\emph{arXiv preprint arXiv:1810.04805}}
  (\bibinfo{year}{2018}).
\newblock


\bibitem[\protect\citeauthoryear{Hashemi, Zamani, and Croft}{Hashemi
  et~al\mbox{.}}{2020}]%
        {hashemi2020guided}
\bibfield{author}{\bibinfo{person}{Helia Hashemi}, \bibinfo{person}{Hamed
  Zamani}, {and} \bibinfo{person}{W~Bruce Croft}.}
  \bibinfo{year}{2020}\natexlab{}.
\newblock \showarticletitle{Guided Transformer: Leveraging Multiple External
  Sources for Representation Learning in Conversational Search}. In
  \bibinfo{booktitle}{\emph{Proceedings of the 43rd International ACM SIGIR
  Conference on Research and Development in Information Retrieval}}.
  \bibinfo{pages}{1131--1140}.
\newblock


\bibitem[\protect\citeauthoryear{Karimzadehgan and Zhai}{Karimzadehgan and
  Zhai}{2011}]%
        {karimzadehgan2011improving}
\bibfield{author}{\bibinfo{person}{Maryam Karimzadehgan} {and}
  \bibinfo{person}{ChengXiang Zhai}.} \bibinfo{year}{2011}\natexlab{}.
\newblock \showarticletitle{Improving retrieval accuracy of difficult queries
  through generalizing negative document language models}. In
  \bibinfo{booktitle}{\emph{Proceedings of the 20th ACM international
  conference on Information and knowledge management}}.
  \bibinfo{pages}{27--36}.
\newblock


\bibitem[\protect\citeauthoryear{Kingma and Ba}{Kingma and Ba}{2014}]%
        {kingma2014adam}
\bibfield{author}{\bibinfo{person}{Diederik~P Kingma} {and}
  \bibinfo{person}{Jimmy Ba}.} \bibinfo{year}{2014}\natexlab{}.
\newblock \showarticletitle{Adam: A method for stochastic optimization}.
\newblock \bibinfo{journal}{\emph{arXiv preprint arXiv:1412.6980}}
  (\bibinfo{year}{2014}).
\newblock


\bibitem[\protect\citeauthoryear{Oddy}{Oddy}{1977}]%
        {oddy1977information}
\bibfield{author}{\bibinfo{person}{Robert~N Oddy}.}
  \bibinfo{year}{1977}\natexlab{}.
\newblock \showarticletitle{Information retrieval through man-machine
  dialogue}.
\newblock \bibinfo{journal}{\emph{Journal of documentation}}
  (\bibinfo{year}{1977}).
\newblock


\bibitem[\protect\citeauthoryear{Peltonen, Strahl, and Flor{\'e}en}{Peltonen
  et~al\mbox{.}}{2017}]%
        {peltonen2017negative}
\bibfield{author}{\bibinfo{person}{Jaakko Peltonen}, \bibinfo{person}{Jonathan
  Strahl}, {and} \bibinfo{person}{Patrik Flor{\'e}en}.}
  \bibinfo{year}{2017}\natexlab{}.
\newblock \showarticletitle{Negative relevance feedback for exploratory search
  with visual interactive intent modeling}. In
  \bibinfo{booktitle}{\emph{Proceedings of the 22nd International Conference on
  Intelligent User Interfaces}}. \bibinfo{pages}{149--159}.
\newblock


\bibitem[\protect\citeauthoryear{Ponte and Croft}{Ponte and Croft}{1998}]%
        {ponte1998language}
\bibfield{author}{\bibinfo{person}{Jay~M Ponte} {and} \bibinfo{person}{W~Bruce
  Croft}.} \bibinfo{year}{1998}\natexlab{}.
\newblock \showarticletitle{A language modeling approach to information
  retrieval}. In \bibinfo{booktitle}{\emph{Proceedings of the 21st annual
  international ACM SIGIR conference on Research and development in information
  retrieval}}. \bibinfo{pages}{275--281}.
\newblock


\bibitem[\protect\citeauthoryear{Qu, Yang, Chen, Qiu, Croft, and Iyyer}{Qu
  et~al\mbox{.}}{2020}]%
        {qu2020open}
\bibfield{author}{\bibinfo{person}{Chen Qu}, \bibinfo{person}{Liu Yang},
  \bibinfo{person}{Cen Chen}, \bibinfo{person}{Minghui Qiu},
  \bibinfo{person}{W~Bruce Croft}, {and} \bibinfo{person}{Mohit Iyyer}.}
  \bibinfo{year}{2020}\natexlab{}.
\newblock \showarticletitle{Open-retrieval conversational question answering}.
  In \bibinfo{booktitle}{\emph{Proceedings of the 43rd International ACM SIGIR
  Conference on Research and Development in Information Retrieval}}.
  \bibinfo{pages}{539--548}.
\newblock


\bibitem[\protect\citeauthoryear{Radlinski and Craswell}{Radlinski and
  Craswell}{2017}]%
        {radlinski2017theoretical}
\bibfield{author}{\bibinfo{person}{Filip Radlinski} {and} \bibinfo{person}{Nick
  Craswell}.} \bibinfo{year}{2017}\natexlab{}.
\newblock \showarticletitle{A theoretical framework for conversational search}.
  In \bibinfo{booktitle}{\emph{Proceedings of the 2017 conference on conference
  human information interaction and retrieval}}. \bibinfo{pages}{117--126}.
\newblock


\bibitem[\protect\citeauthoryear{Rao and Daum{\'e}~III}{Rao and
  Daum{\'e}~III}{2018}]%
        {rao2018learning}
\bibfield{author}{\bibinfo{person}{Sudha Rao} {and} \bibinfo{person}{Hal
  Daum{\'e}~III}.} \bibinfo{year}{2018}\natexlab{}.
\newblock \showarticletitle{Learning to ask good questions: Ranking
  clarification questions using neural expected value of perfect information}.
\newblock \bibinfo{journal}{\emph{arXiv preprint arXiv:1805.04655}}
  (\bibinfo{year}{2018}).
\newblock


\bibitem[\protect\citeauthoryear{Rao and Daum{\'e}~III}{Rao and
  Daum{\'e}~III}{2019}]%
        {rao2019answer}
\bibfield{author}{\bibinfo{person}{Sudha Rao} {and} \bibinfo{person}{Hal
  Daum{\'e}~III}.} \bibinfo{year}{2019}\natexlab{}.
\newblock \showarticletitle{Answer-based adversarial training for generating
  clarification questions}.
\newblock \bibinfo{journal}{\emph{arXiv preprint arXiv:1904.02281}}
  (\bibinfo{year}{2019}).
\newblock


\bibitem[\protect\citeauthoryear{Reddy, Chen, and Manning}{Reddy
  et~al\mbox{.}}{2019}]%
        {reddy2019coqa}
\bibfield{author}{\bibinfo{person}{Siva Reddy}, \bibinfo{person}{Danqi Chen},
  {and} \bibinfo{person}{Christopher~D Manning}.}
  \bibinfo{year}{2019}\natexlab{}.
\newblock \showarticletitle{Coqa: A conversational question answering
  challenge}.
\newblock \bibinfo{journal}{\emph{Transactions of the Association for
  Computational Linguistics}}  \bibinfo{volume}{7} (\bibinfo{year}{2019}),
  \bibinfo{pages}{249--266}.
\newblock


\bibitem[\protect\citeauthoryear{Rocchio}{Rocchio}{1971}]%
        {rocchio1971relevance}
\bibfield{author}{\bibinfo{person}{Joseph Rocchio}.}
  \bibinfo{year}{1971}\natexlab{}.
\newblock \showarticletitle{Relevance feedback in information retrieval}.
\newblock \bibinfo{journal}{\emph{The Smart retrieval system-experiments in
  automatic document processing}} (\bibinfo{year}{1971}),
  \bibinfo{pages}{313--323}.
\newblock


\bibitem[\protect\citeauthoryear{Smucker, Allan, and Carterette}{Smucker
  et~al\mbox{.}}{2007}]%
        {smucker2007comparison}
\bibfield{author}{\bibinfo{person}{Mark~D Smucker}, \bibinfo{person}{James
  Allan}, {and} \bibinfo{person}{Ben Carterette}.}
  \bibinfo{year}{2007}\natexlab{}.
\newblock \showarticletitle{A comparison of statistical significance tests for
  information retrieval evaluation}. In \bibinfo{booktitle}{\emph{Proceedings
  of the sixteenth ACM conference on Conference on information and knowledge
  management}}. \bibinfo{pages}{623--632}.
\newblock


\bibitem[\protect\citeauthoryear{Spina, Trippas, Cavedon, and Sanderson}{Spina
  et~al\mbox{.}}{2017}]%
        {spina2017extracting}
\bibfield{author}{\bibinfo{person}{Damiano Spina}, \bibinfo{person}{Johanne~R
  Trippas}, \bibinfo{person}{Lawrence Cavedon}, {and} \bibinfo{person}{Mark
  Sanderson}.} \bibinfo{year}{2017}\natexlab{}.
\newblock \showarticletitle{Extracting audio summaries to support effective
  spoken document search}.
\newblock \bibinfo{journal}{\emph{Journal of the Association for Information
  Science and Technology}} \bibinfo{volume}{68}, \bibinfo{number}{9}
  (\bibinfo{year}{2017}), \bibinfo{pages}{2101--2115}.
\newblock


\bibitem[\protect\citeauthoryear{Sun and Zhang}{Sun and Zhang}{2018}]%
        {sun2018conversational}
\bibfield{author}{\bibinfo{person}{Yueming Sun} {and} \bibinfo{person}{Yi
  Zhang}.} \bibinfo{year}{2018}\natexlab{}.
\newblock \showarticletitle{Conversational recommender system}. In
  \bibinfo{booktitle}{\emph{The 41st international acm sigir conference on
  research \& development in information retrieval}}.
  \bibinfo{pages}{235--244}.
\newblock


\bibitem[\protect\citeauthoryear{Trippas, Spina, Cavedon, Joho, and
  Sanderson}{Trippas et~al\mbox{.}}{2018}]%
        {trippas2018informing}
\bibfield{author}{\bibinfo{person}{Johanne~R Trippas}, \bibinfo{person}{Damiano
  Spina}, \bibinfo{person}{Lawrence Cavedon}, \bibinfo{person}{Hideo Joho},
  {and} \bibinfo{person}{Mark Sanderson}.} \bibinfo{year}{2018}\natexlab{}.
\newblock \showarticletitle{Informing the design of spoken conversational
  search: Perspective paper}. In \bibinfo{booktitle}{\emph{Proceedings of the
  2018 Conference on Human Information Interaction \& Retrieval}}.
  \bibinfo{pages}{32--41}.
\newblock


\bibitem[\protect\citeauthoryear{Vtyurina, Savenkov, Agichtein, and
  Clarke}{Vtyurina et~al\mbox{.}}{2017}]%
        {vtyurina2017exploring}
\bibfield{author}{\bibinfo{person}{Alexandra Vtyurina}, \bibinfo{person}{Denis
  Savenkov}, \bibinfo{person}{Eugene Agichtein}, {and}
  \bibinfo{person}{Charles~LA Clarke}.} \bibinfo{year}{2017}\natexlab{}.
\newblock \showarticletitle{Exploring conversational search with humans,
  assistants, and wizards}. In \bibinfo{booktitle}{\emph{Proceedings of the
  2017 chi conference extended abstracts on human factors in computing
  systems}}. \bibinfo{pages}{2187--2193}.
\newblock


\bibitem[\protect\citeauthoryear{Wang, Fang, and Zhai}{Wang
  et~al\mbox{.}}{2007}]%
        {wang2007improve}
\bibfield{author}{\bibinfo{person}{Xuanhui Wang}, \bibinfo{person}{Hui Fang},
  {and} \bibinfo{person}{ChengXiang Zhai}.} \bibinfo{year}{2007}\natexlab{}.
\newblock \showarticletitle{Improve retrieval accuracy for difficult queries
  using negative feedback}. In \bibinfo{booktitle}{\emph{Proceedings of the
  sixteenth ACM conference on Conference on information and knowledge
  management}}. \bibinfo{pages}{991--994}.
\newblock


\bibitem[\protect\citeauthoryear{Wang, Fang, and Zhai}{Wang
  et~al\mbox{.}}{2008}]%
        {wang2008study}
\bibfield{author}{\bibinfo{person}{Xuanhui Wang}, \bibinfo{person}{Hui Fang},
  {and} \bibinfo{person}{ChengXiang Zhai}.} \bibinfo{year}{2008}\natexlab{}.
\newblock \showarticletitle{A study of methods for negative relevance
  feedback}. In \bibinfo{booktitle}{\emph{Proceedings of the 31st annual
  international ACM SIGIR conference on Research and development in information
  retrieval}}. \bibinfo{pages}{219--226}.
\newblock


\bibitem[\protect\citeauthoryear{Wang, Liu, Huang, and Nie}{Wang
  et~al\mbox{.}}{2018}]%
        {wang2018learning}
\bibfield{author}{\bibinfo{person}{Yansen Wang}, \bibinfo{person}{Chenyi Liu},
  \bibinfo{person}{Minlie Huang}, {and} \bibinfo{person}{Liqiang Nie}.}
  \bibinfo{year}{2018}\natexlab{}.
\newblock \showarticletitle{Learning to ask questions in open-domain
  conversational systems with typed decoders}. In
  \bibinfo{booktitle}{\emph{Proceedings of the 56th Annual Meeting of the
  Association for Computational Linguistics}}. \bibinfo{pages}{2193--2203}.
\newblock


\bibitem[\protect\citeauthoryear{Wang and Ai}{Wang and Ai}{2021}]%
        {wang2021controlling}
\bibfield{author}{\bibinfo{person}{Zhenduo Wang} {and} \bibinfo{person}{Qingyao
  Ai}.} \bibinfo{year}{2021}\natexlab{}.
\newblock \showarticletitle{Controlling the Risk of Conversational Search via
  Reinforcement Learning}.
\newblock \bibinfo{journal}{\emph{arXiv preprint arXiv:2101.06327}}
  (\bibinfo{year}{2021}).
\newblock


\bibitem[\protect\citeauthoryear{Xu, Wang, Tang, Duan, Yang, Zeng, Zhou, and
  Xu}{Xu et~al\mbox{.}}{2019}]%
        {xu2019asking}
\bibfield{author}{\bibinfo{person}{Jingjing Xu}, \bibinfo{person}{Yuechen
  Wang}, \bibinfo{person}{Duyu Tang}, \bibinfo{person}{Nan Duan},
  \bibinfo{person}{Pengcheng Yang}, \bibinfo{person}{Qi Zeng},
  \bibinfo{person}{Ming Zhou}, {and} \bibinfo{person}{SUN Xu}.}
  \bibinfo{year}{2019}\natexlab{}.
\newblock \showarticletitle{Asking clarification questions in knowledge-based
  question answering}. In \bibinfo{booktitle}{\emph{Proceedings of the 2019
  Conference on Empirical Methods in Natural Language Processing and the 9th
  International Joint Conference on Natural Language Processing
  (EMNLP-IJCNLP)}}. \bibinfo{pages}{1618--1629}.
\newblock


\bibitem[\protect\citeauthoryear{Yang, Qiu, Qu, Guo, Zhang, Croft, Huang, and
  Chen}{Yang et~al\mbox{.}}{2018}]%
        {yang2018response}
\bibfield{author}{\bibinfo{person}{Liu Yang}, \bibinfo{person}{Minghui Qiu},
  \bibinfo{person}{Chen Qu}, \bibinfo{person}{Jiafeng Guo},
  \bibinfo{person}{Yongfeng Zhang}, \bibinfo{person}{W~Bruce Croft},
  \bibinfo{person}{Jun Huang}, {and} \bibinfo{person}{Haiqing Chen}.}
  \bibinfo{year}{2018}\natexlab{}.
\newblock \showarticletitle{Response ranking with deep matching networks and
  external knowledge in information-seeking conversation systems}. In
  \bibinfo{booktitle}{\emph{The 41st international acm sigir conference on
  research \& development in information retrieval}}.
  \bibinfo{pages}{245--254}.
\newblock


\bibitem[\protect\citeauthoryear{Zagheli, Ariannezhad, and Shakery}{Zagheli
  et~al\mbox{.}}{2017}]%
        {zagheli2017negative}
\bibfield{author}{\bibinfo{person}{Hossein~Rahmatizadeh Zagheli},
  \bibinfo{person}{Mozhdeh Ariannezhad}, {and} \bibinfo{person}{Azadeh
  Shakery}.} \bibinfo{year}{2017}\natexlab{}.
\newblock \showarticletitle{Negative feedback in the language modeling
  framework for text recommendation}. In \bibinfo{booktitle}{\emph{European
  Conference on Information Retrieval}}. Springer, \bibinfo{pages}{662--668}.
\newblock


\bibitem[\protect\citeauthoryear{Zamani, Dumais, Craswell, Bennett, and
  Lueck}{Zamani et~al\mbox{.}}{2020}]%
        {zamani2020generating}
\bibfield{author}{\bibinfo{person}{Hamed Zamani}, \bibinfo{person}{Susan
  Dumais}, \bibinfo{person}{Nick Craswell}, \bibinfo{person}{Paul Bennett},
  {and} \bibinfo{person}{Gord Lueck}.} \bibinfo{year}{2020}\natexlab{}.
\newblock \showarticletitle{Generating clarifying questions for information
  retrieval}. In \bibinfo{booktitle}{\emph{Proceedings of The Web Conference
  2020}}. \bibinfo{pages}{418--428}.
\newblock


\bibitem[\protect\citeauthoryear{Zhang, Chen, Ai, Yang, and Croft}{Zhang
  et~al\mbox{.}}{2018}]%
        {zhang2018towards}
\bibfield{author}{\bibinfo{person}{Yongfeng Zhang}, \bibinfo{person}{Xu Chen},
  \bibinfo{person}{Qingyao Ai}, \bibinfo{person}{Liu Yang}, {and}
  \bibinfo{person}{W~Bruce Croft}.} \bibinfo{year}{2018}\natexlab{}.
\newblock \showarticletitle{Towards conversational search and recommendation:
  System ask, user respond}. In \bibinfo{booktitle}{\emph{Proceedings of the
  27th acm international conference on information and knowledge management}}.
  \bibinfo{pages}{177--186}.
\newblock


\bibitem[\protect\citeauthoryear{Zhao, Zhang, Ding, Xia, Tang, and Yin}{Zhao
  et~al\mbox{.}}{2018}]%
        {zhao2018recommendations}
\bibfield{author}{\bibinfo{person}{Xiangyu Zhao}, \bibinfo{person}{Liang
  Zhang}, \bibinfo{person}{Zhuoye Ding}, \bibinfo{person}{Long Xia},
  \bibinfo{person}{Jiliang Tang}, {and} \bibinfo{person}{Dawei Yin}.}
  \bibinfo{year}{2018}\natexlab{}.
\newblock \showarticletitle{Recommendations with negative feedback via pairwise
  deep reinforcement learning}. In \bibinfo{booktitle}{\emph{Proceedings of the
  24th ACM SIGKDD International Conference on Knowledge Discovery \& Data
  Mining}}. \bibinfo{pages}{1040--1048}.
\newblock


\end{thebibliography}

\end{document}